\documentclass[%
 reprint,
showpacs,
 amsmath,amssymb,
 aps,
 pra,
longbibliography
]{revtex4-1}

\usepackage{graphicx} 
\usepackage{dcolumn} 
\usepackage{bm} 

\usepackage[utf8]{inputenc}
\usepackage[T1]{fontenc}
\usepackage{times}
\usepackage{flafter} 
\usepackage{textgreek} 
\usepackage{mathrsfs} 
\usepackage[usenames,dvipsnames]{xcolor}
\usepackage[hidelinks,urlcolor=blue, colorlinks=true,citecolor=blue, linkcolor=blue ]{hyperref}

\newcommand{\ii}{ i }
\newcommand{\ee}{\mathrm{e}}

\renewcommand{\Re}{\operatorname{Re}}
\renewcommand{\Im}{\operatorname{Im}}

\newcommand{\fullwf}{\mathit{\Phi}}
\newcommand{\velel}{{\hat v_\mathrm{ee}}}

\newcommand{\pr}{^\prime}

\newcommand{\derivative}{\partial}
\newcommand{\odd}{\text{ odd}}
\newcommand{\even}{\text{ even}}

\newcommand{\Vee}{\hat{v}_\mathrm{ee}}
\newcommand{\Ge}{\hat{\mathit{\Gamma}}_\mathrm{e}}
\newcommand{\tr}{\mathrm{Tr}}
\newcommand{\ket}[1]{| #1 \rangle}
\newcommand{\bra}[1]{\langle #1 |}
\newcommand{\ketbra}[2]{| \vphantom{#2} #1 \rangle\langle \vphantom{#1} #2 |}
\newcommand{\ketketbrabra}[4]{| \vphantom{#2#3#4} #1 \rangle{| \vphantom{#4#1#3} #2 \rangle\langle 
\vphantom{#1#2#4} #3 |}\langle\vphantom{#1#2#3} #4 |}
\renewcommand{\d}[2]{\frac{d #1}{d #2}}
\newcommand{\pd}[2]{\frac{\partial #1}{\partial #2}}
\newcommand{\dd}[2]{\frac{d^2 #1}{d #2^2}}
\newcommand{\com}[2]{\left[#1 \vphantom{#2} , #2 \vphantom{#1}  \right]}
\newcommand{\acom}[2]{\left[#1 \vphantom{#2} , #2 \vphantom{#1}  \right]_{+}}
\providecommand{\mean}[1]{\langle#1\rangle}
\newcommand{\braket}[2]{\langle #1 \vphantom{#2} |  #2 \vphantom{#1} \rangle} 
\newcommand{\fbraket}[3]{\langle #1 \vphantom{#2#3} | #2 | #3 \vphantom{#1#2} \rangle}
\newcommand{\abs}[1]{|#1|}
\newcommand{\pdd}[2]{\frac{\partial^2 #1}{\partial #2^2}}

\newcommand{\iek}[1]{\left(#1\right)}
\newcommand{\kiek}[1]{\left[#1\right]}

\newcommand{\ketrno}[1]{|\tilde{#1} \rangle}

\newcommand{\mat}[1]{\mathbf{\mathcal{#1}}}

\DeclareMathOperator{\Tr}{Tr}

\makeindex

\begin{document}

\title{Strong-field absorption and emission of radiation in two-electron systems calculated with time-dependent natural orbitals}

\author{M.\ Brics}
\author{J.\ Rapp}
\author{D.\ Bauer}%
\affiliation{%
 Institut für Physik, Universität Rostock, 18051 Rostock, Germany
}

\date{\today}

\begin{abstract}
Recently introduced time-dependent renormalized-natural-orbital theory (TDRNOT) is based on the equations of motion for the so-called natural orbitals, i.e., the eigenfunctions of the one-body reduced density matrix. Exact TDRNOT can be formulated for any time-dependent two-electron system in either spin configuration. In this paper, the method is tested against high-order harmonic generation (HHG) and  Fano profiles in absorption spectra with the help of a  numerically exactly solvable one-dimensional model He atom, starting from the spin-singlet ground state. Such benchmarks are challenging because Fano profiles originate from transitions involving autoionizing states, and HHG is a strong-field phenomenon well beyond linear response. TDRNOT with just one natural orbital per spin in the helium spin-singlet case is equivalent to time-dependent Hartree-Fock or  time-dependent density functional theory (TDDFT) in  exact exchange-only approximation.  It is not unexpected that TDDFT fails in reproducing Fano profiles due to the lack of doubly excited, autoionizing states. HHG spectra, on the other hand, are widely believed to be well-captured by TDDFT. However, HHG spectra of helium may display a second plateau that originates from simultaneous HHG in He$^+$ and neutral He. It is found that already TDRNOT with two natural orbitals per spin is sufficient to capture this effect as well as the Fano profiles on a qualitative level. With more natural orbitals (6--8 per spin) quantitative agreement can be reached. Errors due to the truncation to a finite number of orbitals are identified. 
\end{abstract}


\maketitle

\section{Introduction}
\label{sec:intro} 
Time-dependent few and many-body methods for driven quantum dynamics beyond linear response are urgently needed both to study fundamental effects and for possible technological applications (involving strong light fields, for instance). Depending on the system studied, the ``exact'' solution would involve the numerical solution of the time-dependent Schr\"odinger, Pauli, Klein-Gordon, or Dirac equation, possibly with quantized electromagnetic field. Unfortunately, this is---in full dimensionality and with strong, long-wavelength lasers---even in the simplest case of the time-dependent Schr\"odinger equation (TDSE) possible only for (at most) two particles.

The most widely used and favorably scaling approach to electronic structure problems is density functional theory \cite{KohnNobel1999, Dreizler1990}. Its time-dependent version, time-dependent density functional theory (TDDFT) \cite{UllrichBook, Ullrich2013}, often misses important correlation effects \cite{florian2, rabi-1}. Other, more systematic approaches such as multi-configurational time-dependent Hartree-Fock (MCTDHF) \cite{Zanghellini2004, Hochstuhl2010} and variants of it \cite{Sato2013,Sato2015} or  time-dependent configuration interaction (TDCI) \cite{Greenman2010, Pabst2013, Karamatskou2014} and related approaches \cite{Hochstuhl2012, Bauch2014} are (much) more demanding but capture correlation effects better \cite{Ishikawa2015,Hochstuhl2014}. Recently, we have introduced  time-dependent renormalized-natural-orbital theory (TDRNOT) \cite{tdrnot, tdrnot2, tdrnot3}, which is based on equations of motion for the so-called natural orbitals (NOs), i.e., the eigenfunctions of the one-body reduced density matrix (1RDM) \cite{Loewdin1955, Coleman2000, AppelPhDThesis,GiesbertzPhDThesis,Sato2014}. While the proof that the natural orbitals form the best basis is restricted to two electrons \cite{Giesbertz2014}, the educated guess (and hope) is that TDRNOT calculations with a very limited number of NOs allow to well surpass TDDFT with relatively little computational overhead. In recent papers, we have demonstrated already that TDRNOT performs well in treating phenomena where TDDFT with known exchange-correlation potentials fail, e.g., autoionization \cite{tdrnot}, Rabi flopping \cite{tdrnot2}, and nonsequential ionization \cite{tdrnot3}. In this work, we will continue along the same line by focusing on the emission and absorption of radiation. In fact, high-harmonic generation (HHG) and absorption spectroscopy (AS) are of eminent importance in stong-field laser physics. HHG is the basis of ``attosecond science'' \cite{Krausz2009,SchultzVrakking2013} while  transient AS provides all-optical means to follow correlated processes in matter \cite{Ott2013, Liu2015}.  

The paper is organized as follows.  The theoretical methods and the way to calculate spectra are described in Sec.~\ref{sec:theor}. In Sec.~\ref{sec:res} we benchmark the performance
of TDRNOT on HHG spectra and Fano profiles, before we conclude and give an outlook in Sec.~\ref{sec:concl}. Some of the detailed derivations are given in Appendices \ref{sec:deriv_EOM} and \ref{sec:phase_conv}.

Atomic units (a.u.) are used throughout unless noted otherwise.

\section{Theoretical methods}
\label{sec:theor}
By numerically solving the  time-dependent Schr\"odinger equation (TDSE) for a one-dimensional helium atom in the laser field exactly, we obtain a reference result for the corresponding  TDRNOT calculation involving $N$ NOs. With increasing $N$ the TDRNOT spectra should converge to the TDSE results, as the TDRNOT equations of motion (EOM) are exact for two electrons. In this Section the model atom, the EOM, and the method to calculate the absorption spectra are introduced. 

\subsection{Model helium atom}

The Hamiltonian of the widely used \cite{Eberly, Haan1994, Dieter97, Lappas_Leeuwen,Lein2005, florian, florian2, rabi-1} one-dimensional helium model atom is
\begin{align}
  \hat H^{(1, 2)}(t)
    &=
      \hat h^{(1)}(t)
      + \hat h^{(2)}(t)
      + \velel^{(1, 2)}-\ii\Ge^{(1)}-\ii\Ge^{(2)} \label{hamiltonian}
\end{align}
where upper indices indicate the action on either electron $\mathrm 1$, electron $2$, or both. The single-particle Hamiltonian reads $\hat 
h(t)=\hat{h}_{\mathrm{A}}+\hat{h}_{\mathrm{L}}(t)$  with
\begin{align}
  \hat{h}_{\mathrm{A}} &=
      \frac{\hat p^2}{2}
      - \frac{2}{\sqrt{\hat x^2 + \varepsilon_{\mathrm{ne}}}},
\end{align}
\begin{align}
\hat{h}_{\mathrm{L}}(t) &=E(t)\hat x
\end{align}
 (dipole approximation and length gauge),
the electron-electron interaction 
\begin{align}
  \velel^{(1, 2)}
    &=
      \frac{1}{\sqrt{\left(\hat x^{(1)}-\hat x^{(2)}\right)^2 + \varepsilon_\mathrm{ee}}},
\end{align}
and $-i\Ge$ is an imaginary potential to absorb outgoing electron flux.
The electron-ion smoothing parameter $\varepsilon_\mathrm{ne}=0.50$ 
is chosen such that the groundstate energy of  $\mathrm{He}^+$,  $E_0^\mathrm{He^+}=-2.0$, is recovered. The electron-electron smoothing parameter 
$\varepsilon_\mathrm{ee}=0.33$ is 
tuned to yield  the neutral-He energy $E_0^\mathrm{He}=-2.9$. 

\subsection{Equations of motion (EOM)}
For a helium atom, the TDSE
\begin{equation}
\ii \pd{}{t} \ket{\fullwf (t)} =\hat H^{(1, 2)}(t)\ket{\fullwf (t)}
\label{eq:TDSE}
\end{equation}
describes the time evolution of the two-electron
state $\ket{\fullwf (t)}$.
The starting point for TDRNOT in the two-electron case is the pure 2-body density matrix(2DM)
\begin{align}
  \hat\gamma_2(t)
    &=
\ketbra{\fullwf(t)}{\fullwf(t)}. \label{eq:2dm}
\end{align}
From the TDSE~\eqref{eq:TDSE} the EOM for the 2DM
\begin{equation}
\begin{split}
 \ii \dot{\hat \gamma}_2(t) =& \com{\hat h^{(1)}(t)
      + \hat h^{(2)}(t) 
      + \velel^{(1, 2)}}{\hat \gamma_2(t)}\\
&-\ii\acom{\Ge^{(1)}+\Ge^{(2)}}{\hat \gamma_2(t)}
 \label{eq:EOMg2}
\end{split}
\end{equation}
results.
The 1RDM $\hat\gamma_1(t)$ reads
\begin{align}
  \hat\gamma_1(t)
    &=
      \sum_{i=1}^2\Tr_i
      \hat\gamma_2(t)
    =
      2\Tr_1
      \hat\gamma_2(t)
    =
      2\Tr_2
      \hat\gamma_2(t), \label{eq:def-1rdm}
\end{align}
where the partial trace $\Tr_i$ means tracing out all degrees of freedom of particle $i$. The EOM for $\hat \gamma_1$ can be derived by taking the time derivative of \eqref{eq:def-1rdm},
\begin{equation}
\begin{split} 
 \ii \dot {\hat \gamma}_1(t) =& \com{\hat h(t)}{\hat\gamma_{1}(t)}
  + 2\tr_{2} \com{\Vee}{\hat\gamma_{2}(t)} -\ii \acom{\Ge}{\hat\gamma_{1}(t)}\\ &-2\ii 
\tr_{2} \acom{\Ge^{(2)}}{ \hat \gamma_{2}(t)}.
\label{eq:eom_g1}
\end{split}
\end{equation}
The NOs $|k(t)\rangle$ and occupation numbers (ONs) $n_k(t)$ are defined as eigenstates and eigenvalues of the 1RDM, 
respectively,
\begin{align}
  \hat\gamma_1(t)
  \ket{k(t)}
    &=
      n_k(t)
      |k(t)\rangle.
\end{align}
As $\hat\gamma_1(t)$ is hermitian, the ONs $n_k(t)$ are real and the NOs $|k(t)\rangle$ form an orthonormal complete 
basis set. Renormalized NOs (RNOs) are defined as
\begin{equation}
\ket{\tilde k(t)}=\sqrt{n_k(t)} \ket{k(t)}
\end{equation}
so that
\begin{equation}
 \hat\gamma_1(t) = \sum_k \ketbra{\tilde k (t)}{\tilde k (t)}, 
\label{eq:rno_def}
\end{equation}
and $ \hat\gamma_2(t)$ can be expanded in RNOs as
\begin{align}
  \hat\gamma_2(t)
    &=
      \sum_{ijkl}
      \tilde \gamma_{2,ijkl}(t)
      \ketketbrabra{\tilde i(t)}{\tilde j(t)}{\tilde k(t)}{\tilde l(t)}. \label{eq:formal-2dm-expansion}
\end{align}

The EOM for the RNOs  read
\begin{equation}
\begin{split} 
  \ii\derivative_t\ketrno{n}
    &=
      \iek{\hat{h}(t)-\ii\Ge}\ketrno{n}+ {\mat{A}}_n(t) \ketrno{n}\\
       &+ \sum_{k\neq n} {\mat{B}}_{nk}(t)\ketrno{k}
       + \sum_k {\mat{\hat{C}}}_{nk} (t) \ketrno{k}, \label{eq:eom-rno}
\end{split}
\end{equation}
with 
\begin{equation}
\begin{split} 
  \mat{{A}}_n(t)
    &=
      -\frac{1}{{n}_n(t)}\Re\sum_{jkl}
      \tilde\gamma_{2,njkl}(t)
      \langle \tilde k \tilde l |\velel|\tilde n \tilde j \rangle \\
      &-\frac{1}{2{n}_n(t)}\iek{\fbraket{\tilde n}{\hat{h}(t)}{\tilde n}-\fbraket{\tilde n'}{\hat{h}(t)}{\tilde n'}}\\
&-2\ii  \sum_{jl} \tilde 
\gamma_{2,njnl}(t) \fbraket{\tilde l}{\Ge}{\tilde j} ,
\end{split} 
\end{equation}
\begin{align}
\begin{split}
 &{\mat{B}}_{nk}(t)
    =
      \frac{2}{{n}_k(t)-{n}_n(t)}\sum_{jpl}\Bigl[
        \tilde \gamma_{2,kjpl}(t)
        \langle \tilde p \tilde l |\velel| \tilde n\tilde j \rangle \Bigl.\\
        & -\Bigr. \tilde\gamma_{2,plnj}(t)
        \langle \tilde k \tilde j |\velel| \tilde p \tilde l \rangle
      \Bigr]
-2\ii\frac{1}{n_n(t)-n_k(t)}\fbraket{\tilde k}{\Ge}{\tilde n}  \\
&-4\ii\frac{n_n(t)}{n_n(t)-n_k(t)}\sum_{jl}\tilde \gamma_{2, kjnl}(t) \fbraket{\tilde l}{\Ge}{\tilde{j}} , \quad n_k(t) \neq n_n(t)
\label{eq:Bnk1}
\end{split}
\end{align}
%
%
%
%
\begin{align}
 {\mat{\hat{C}}}_{nk}(t)
    &=
      2\sum_{jl} \tilde \gamma_{2,kjnl}(t)\langle \tilde l|\velel| \tilde j \rangle, \label{eq:eom-end} 
\end{align}
and the ``prime operator'' acting on a positive integer $k$ as
\begin{align}
  k\pr
    &=
      \begin{cases}
        k+1 & \text{if $k\odd$}\\
        k-1 & \text{if $k\even$},
      \end{cases}
  &
  k
    &>
      0. \label{eq:prime-operator}
\end{align}
For derivations see Appendices \ref{sec:deriv_EOM} and \ref{sec:phase_conv}. Also, note that the time argument of the RNOs is suppressed, and we can apply \eqref{eq:Bnk1} only if at time $t$ ON $n_k(t) \neq n_n(t)$ (otherwise see Appendix~\ref{sec:deriv_EOM}).

\subsection{Absorption  spectra}
\label{sec:spectra} 
Let  the classical electric field of a laser be polarized in $x$~direction and propagating in $y$~direction, $E_{\mathrm{in}}(t-y/c)=E_{\mathrm{in}}(t-\alpha y)$ and the atom be placed at the 
origin. The laser field $E_{\mathrm{in}}$ induces a dipole, and the atom responds, generating a field $E_{\mathrm{gen}}(y, 
t)$ also polarized in 
$x$ direction. 
The spectral distribution $S(\omega)$ reads
\begin{equation}
  S(\omega)=\abs{E(\omega)}^2=\frac{1}{2\pi}\left| \int_{-\infty}^{+\infty} \!\!\!\!  dt \, E(y, t) \ee^{\ii \omega t}\right|^2 
\label{eq:spectrum}
\end{equation}
where $E(y, t)$ is the total field and $E(\omega)$ its Fourier transform at position $y$. 
The total electric field  $E(y, t)$ is determined by the wave equation in propagation direction with a polarization term as a source \cite{Baggesen2011}, 
\begin{equation}
\iek{\pdd{}{y}-\frac{1}{c^2}\pdd{}{t}} E(y, t)=\frac{1}{\epsilon_{0}c^2}\pdd{}{t}\mean{d(t)}\delta(y) 
\label{eq:Maxwell}
\end{equation}
where $\mean{d(t)}$ is the expectation value of the $x$ component of the atomic dipole, which is the quantum mechanical single-atom input obtained from the one-dimensional helium model. The dipole is $\mean{d(t)}=-\sum_i\mean{x_i(t)}$ with 
$\mean{x_i(t)}=\fbraket{\fullwf(t)}{\hat x^{(i)}}{\fullwf(t)}$.   

The relevant solution of \eqref{eq:Maxwell} consists of an incoming and two counter propagating waves, generated by the
 induced dipole in the atom,
\begin{equation}
\begin{split}
 &E(y, t)=E_{\mathrm{in}}(t-\alpha y)\\&-2\pi\alpha\kiek{\theta(y)\mean{\dot d(t-\alpha y)}+\theta(-y)\mean{\dot d(t+\alpha y)}}
\end{split}
\label{eq:Maxwell_sol}
\end{equation}
where $\theta(\pm y)$ is the Heaviside step function and $\mean{\dot d(t\pm\alpha y)}$ is the expectation value of the dipole velocity. The generated wave traveling in  propagation direction of the incoming pulse $\mean{\dot d(t-\alpha y)}$ may interfere destructively with the latter (absorption). In order to identify what is absorbed and what is emitted, the spectrum of the incoming laser field $E_{\mathrm{in}}(\omega)$ is subtracted from the total one \cite{Baggesen2011},
\begin{equation}
  \begin{split}
   S_{\mathrm{resp}}(\omega)&= S(\omega) - \abs{E_{\mathrm{in}}(\omega)}^2 \\
   &=\frac{4\pi^2 \alpha^2}{\omega^2}\abs{\ddot d (\omega)}^2 +\frac{4\pi\alpha }{\omega}\Im\kiek{E^\ast_{\mathrm{in}}(\omega)\ddot d (\omega)}
  \end{split}
\label{eq:S_abs}
\end{equation}
where $\ddot d(\omega)$ is the Fourier transform of 
the expectation value of the dipole acceleration.  Positive values of  $S_{\mathrm{resp}}(\omega)$ indicate emission, negative values absorption. 

If $\abs{E_{\mathrm{gen}}}\ll \abs{E_{\mathrm{in}}}$ in the frequency range where the incoming laser has components one can 
approximate
\begin{equation}
  S_{\mathrm{resp}}(\omega) \approx S^{\mathrm L}_{\mathrm{resp}}(\omega)= \frac{4\pi \alpha}{\omega}\Im\kiek{E^\ast_{\mathrm{in}}(\omega)\ddot d 
(\omega)} . \label{eq:sl}
\end{equation}
In HHG  we are rather interested in frequencies where the incoming laser has no or negligible components, i.e.,
\begin{equation}
  S_{\mathrm{resp}}(\omega) \approx S^{\mathrm {NL}}_{\mathrm{resp}}(\omega) =\frac{4\pi^2 \alpha^2}{\omega^2}\abs{\ddot d (\omega)}^2 .
\label{eq:snl}
\end{equation}
Note that in the literature one finds similar expressions derived using different approaches leading to, however, different 
pre-factors and $\omega$ scaling. In \cite{Gaarde11,Yang15}, for instance, one finds that $S^{\mathrm L}_{\mathrm{resp}}(\omega) 
\sim \frac{1}{\omega^2}\Im\kiek{E^\ast_{\mathrm{in}}(\omega)\ddot d(\omega)}$. The ``standard way'' to calculate HHG spectra is via  $S^{\mathrm {NL}}_{\mathrm{resp}}(\omega)\sim \abs{\ddot d (\omega)}^2$ (derived from Larmor's formula). On the other hand,  \eqref{eq:snl} is in agreement with quantum-electrodynamical calculations \cite{Diestler08}.

Numerically, it is advantageous to evaluate $\dd{}{t}\mean{d}$ using Ehrenfest's theorem,
\begin{equation}
\begin{split}
  \dd{}{t}\mean{d}&= -2\left\langle\d{}{x_i}\frac{2}{\sqrt{x_i^2 + 
\varepsilon_{\mathrm{ne}}}}\right\rangle-2\dot A(t)\\&=2\left\langle\frac{2x_i}{(x_i^2 + \varepsilon_{\mathrm{ne}})^{3/2}}\right\rangle +2E(t).
\label{eq:acceleration}
\end{split}
\end{equation}

\section{Results}
\label{sec:res}
Results from  TDRNOT calculations for AS and HHG, together with the corresponding TDSE benchmarks, will be presented in this Section. All results were obtained starting from the spin-singlet ground state, which was calculated via imaginary-time propagation. Real-time propagation was performed with enabled imaginary 
potential on a grid with $500$ grid points (in each spatial direction) with a grid spacing of $0.4$. 

\begin{figure}[htbp]
\includegraphics[width=0.99\columnwidth]{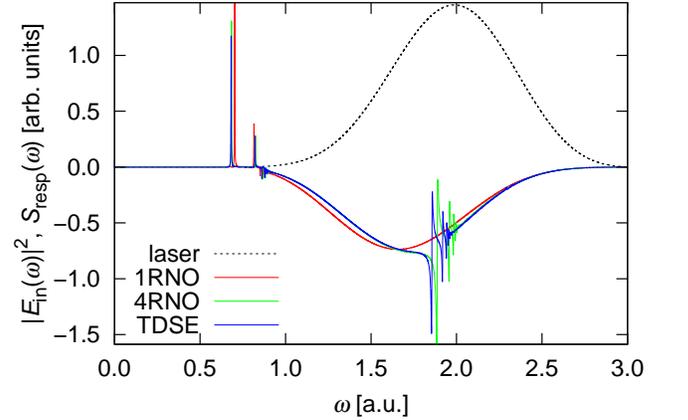}
\caption{(Color online) Absorption spectra calculated using TDSE and TDRNOT with one and four RNOs. The dashed curve indicates the spectrum of the incoming laser field.}
\label{fig:abs_spec}
\end{figure}
\subsection{Fano profiles in absorption spectra}
\label{sec:fano}
A 25-nm ($\omega=1.84$) linearly polarized  $N_{\mathrm{cyc}}=3$-cycle  $\sin^2$-shaped laser pulse of duration $T=2 \pi N_{\mathrm{cyc}}/\omega=10.4=0.25 \; \mathrm {fs}$ was applied to the model helium atom. The vector potential in dipole 
approximation reads
\begin{equation}
  A(t)=A_0 \sin^2\iek{\frac{\omega t}{2 N_{\mathrm{cyc}}}}\sin(\omega t) \quad \text{for } 0\leq t \leq T
\end{equation}
and zero otherwise. The chosen peak intensity $I_0=(\omega A_0)^2$ corresponds to  $I_0=1.0 \times 10^{12} \; \mathrm{W/cm^2}$. In order to obtain a high frequency resolution the quantum propagation was continued for 
$T_{\mathrm{free}}=10000$ after the  laser pulse.  An exponential decay 
$
 W(t)=\ee^{-\beta\theta(t-T) (t-T)}
$
with $\beta=2.5\times 10^{-5}$
was multiplied to $\ddot x(t)$ in order to mimic the decay of excited-state population due to spontaneous emission.

Figure~\ref{fig:abs_spec} shows absorption spectra calculated with TDRNOT using one and four RNOs per spin, together with the exact TDSE benchmark result.  The lines in the frequency interval $[0.6,0.9]$ correspond to 
transitions between the groundstate and singly excited states, the lines lying within $[1.8,2.2]$ to transitions involving doubly excited states where---in a single-particle picture---the lower-energetic electron 
is in the first excited level. Note that the strength of the  lines in the frequency range $[0.6,0.9]$ is sensitive to the choice of the damping coefficient $\beta$. However, for the purpose of benchmarking the TDRNOT results that is not important as long as the same $\beta$ is used for both TDSE and TDRNOT.
\begin{figure}[htbp]
\includegraphics[width=0.99\columnwidth]{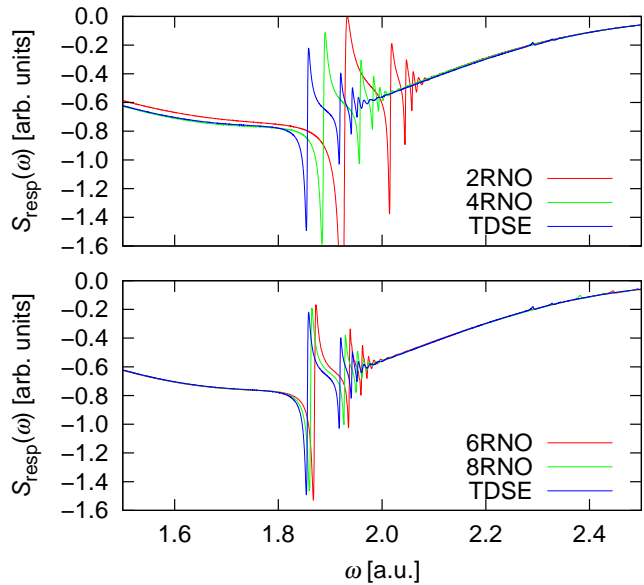}
\caption{(Color online) Fano profiles due to doubly excited states calculated using TDSE and TDRNOT with two, four, six , and eight RNOs. }
\label{fig:fano}
\end{figure}
While the lines involving singly excited states are present already in the 1RNO calculation, the Fano lineshapes  for $\omega\in [1.8,2.2]$ are only reproduced with more than one RNO per spin. This is not surprising, as one RNO per spin is equivalent to time-dependent Hartree-Fock, which is identical to TDDFT in exact-exchange-only approximation. The latter is known to miss doubly excited states \cite{Maitra2004}.

Figure~\ref{fig:fano} shows that already the 2RNO result reproduces  the  Fano profiles but their position  is not yet satisfactory. We find that by adding more RNOs the convergence to the TDSE result is ``quasi-monotonous,'' i.e., it transiently worsens for odd numbers of RNOs per spin but then improves for the subsequent even number. Six RNOs per spin give reasonable spectra if the lowest series of doubly excited states is relevant. We found that 14 RNOs are required if the next series (with the lower electron in the second excited level) is involved.

\begin{figure*}[htbp]
\includegraphics[width=0.99\textwidth]{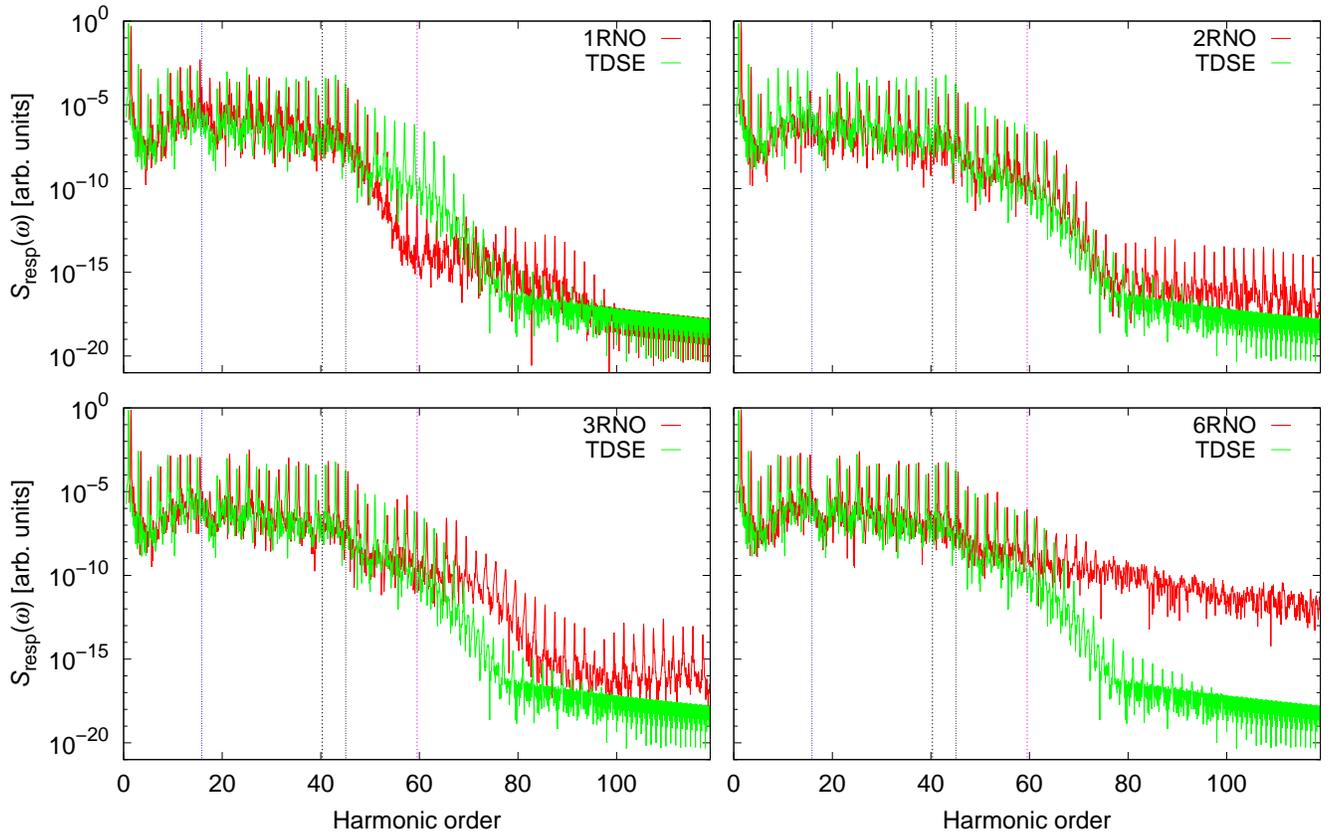}
\caption{(Color online) Comparison of HHG spectra obtained with TDSE and TDRNOT with one, two, three, and six RNOs. For a better comparison the TDRNOT spectra are shifted to the right by half a harmonic order. The vertical lines in each panel indicate (from left to right) the ionization potential  $I_p$ of neutral He, the cut-off $3.17\,U_p+I_p$, the corrected cut-off $3.17\, U_p+1.3\,I_p$ obtained by Lewenstein \cite{Lewenstein}, and the He$^+$ cut-off $3.17\,U_p+I_p^{\mathrm{He}^+}$.}
\label{fig:hhg}
\end{figure*}

\subsection{HHG spectra}
\label{sec:hhg}
One might expect that the calculation of HHG spectra is an easy task, even for TDDFT in the simple exact-exchange-only approximation. However, apart from possible correlation effects one has to keep in mind that the celebrated HHG cut-off $3.17\,U_p + I_p$ (with $I_p=|E_0|$ the ionization potential and $U_p=A_0^2/4$ the ponderomotive energy) \cite{Lewenstein} involves $I_p$. Hence, a multiple step-like structure may arise  because of the different $I_p$ for different charge states in multi-electron systems.

In this HHG part of our work we applied a rather long ($N_{\mathrm{cyc}}= 15$), flat-top 800-nm ($\omega=0.057$) pulse in order to generate well-defined, sharp harmonic lines. The up and down ramping was $\sin^2$-shaped over two cycles.  The peak intensity of the laser pulse was $I_0=1.0 \times 10^{14} \; \mathrm{W/cm^2}$.

%
%

Figure~\ref{fig:hhg} shows the HHG spectra obtained with TDRNOT using one, two, three, and six RNOs, together with the TDSE benchmark. The 1RNO result (i.e., TDHF or exact-exchange-only TDDFT) misses the proper extension of the plateau due to HHG in $\mathrm{He^+}$. Instead, it gives an unphysical second plateau, which is a replica of the first plateau due to the nonlinearities in the EOM. Note that the laser parameters are not in the regime where a second plateau due to single-photon double recombination occurs \cite{Koval07}.  

\begin{figure}[htbp]
\includegraphics[width=0.99\columnwidth]{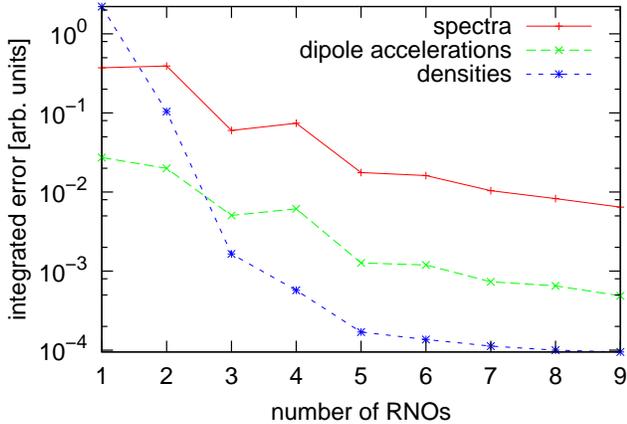}
\caption{(Color online) Integrated error for HHG spectra,  dipole accelerations $\mean{\ddot d}$, and densities in the TDRNOT results for one up to nine RNOs per spin.}  
\label{fig:error}
\end{figure}

The 2RNO spectrum in Fig.~\ref{fig:hhg} reproduces the 
 $\mathrm{He^+}$ cut-off very well. However, unphysical harmonic peaks emerge on a level of $10^{-13}$ well above even the 100th harmonic. The situation seems to worsen if more RNOs are added. The reason for these unphysical high harmonics is truncation. By adding more and more RNOs, TDRNOT is able to describe transitions to more and more doubly excited states. However, the energetic position of each new series of doubly excited states is never correct in the first place. It only converges to the right position if one adds even more RNOs (which bring new, initially poorly positioned series). These wrong states are responsible for the unphysical harmonics. Nevertheless the quantitative agreement improves with increasing number of RNOs because the harmonics up to the  $\mathrm{He^+}$ cut-off match better the TDSE benchmark result. In fact, for 6 RNOs per spin the TDRNOT and TDSE HHG peaks up to the 63rd harmonic agree very well. Figure~\ref{fig:error} shows the error of the TDRNOT HHG spectra, i.e., the square of the difference between the TDRNOT and the TDSE spectrum, integrated over all frequencies. 
It is seen that the error decreases with number of RNOs although not strictly monotonously. Similarly, the errors in the dipole acceleration $\mean{\ddot d}$ (integrated 
over the laser pulse duration) and in the density (integrated over space and the laser pulse duration) were calculated and are included in  Fig.~\ref{fig:error}.

\section{Conclusion}
\label{sec:concl}
In this work, we tested further the recently introduced time-dependent renormalized-natural-orbital theory (TDRNOT). The method is based on the equations of motion for the  renormalized natural orbitals (RNOs), i.e., the time-dependent eigenfunctions  of the one-body reduced density matrix, normalized to their eigenvalues.  TDRNOT was applied again to a numerically exactly solvable model helium atom with the focus on the absorbed and emitted radiation. Both absorption and high-harmonic generation (HHG) spectra are ``simple observables'' in the sense that the single-particle density is sufficient to calculate them. Hence, one could have expected that  time-dependent density functional theory (TDDFT) in exact exchange-only approximation would work well. However, we showed that TDRNOT with just one RNO per spin (which equals TDDFT in exact exchange-only approximation) fails to reproduce Fano line shapes in absorption spectra due to the absence of doubly excited states. Adding more RNOs the Fano line shapes are captured well by TDRNOT. The 1-RNO or TDDFT result for HHG spectra was qualitatively wrong because it lacked the correct cut-off originating from HHG in He$^+$, while predicting an unphysical second plateau. Again, with increasing number of RNOs the agreement with the benchmark result from the exact numerical solution of the time-dependent Schr\"odinger equation improves. However, unphysical HHG peaks at high frequencies emerge because of the necessary truncation of the number of RNOs. 

A TDRNOT implementation for helium in full dimensionality is under way. For more than two particles approximations for  the two-body reduced density matrix expansion coefficients $\tilde \gamma_{2,ijkl}(t)$ in  \eqref{eq:formal-2dm-expansion} are required. Testing such approximations will be the subject of future work.

\section*{Acknowledgment}
This work was supported by the SFB 652 of the German Science Foundation (DFG).

\appendix
\section{Derivation of EOM}
\label{sec:deriv_EOM} Multiplying \eqref{eq:eom_g1} from the right with $\ket{\tilde n}$ and using the 2DM expansion in RNOs~\eqref{eq:formal-2dm-expansion} one obtains
\begin{align}
\begin{split}
 \ii \dot {\hat\gamma}_{1}(t) \ket{\tilde n} 
=&n_n(t)(\hat h(t)-\ii \Ge) \ket{\tilde n} \\
&-\sum_k \fbraket{\tilde k}{\hat h(t)+\ii \Ge}{\tilde n}\ket{\tilde k} \\
&+2 n_n(t)  \sum_{ijk} 
\tilde \gamma_{2,ijnk}(t) \fbraket{\tilde k}{\Vee}{\tilde j} \ket{\tilde i}\\
&-2\sum_{ijkl} \tilde \gamma_{2,ijkl}(t) \fbraket{\tilde k \tilde l}{\Vee}{\tilde n \tilde j} \ket{\tilde i}
\\
&-4\ii n_n(t)  \sum_{ijl}  \tilde \gamma_{2,ijnl}(t) \fbraket{\tilde l}{\Ge}{\tilde j} \ket{\tilde i},
\label{eq:gamma_1}
\end{split}
\end{align}
which multiplied from the left with $\bra{\tilde l}$ leads to
\begin{align}
\begin{split}
 &\ii\fbraket{\tilde l}{ \dot {\hat\gamma}_{1}(t) }{\tilde n} 
=\iek{n_n(t)-n_l(t)}\fbraket{\tilde l}{\hat h(t)}{\tilde n} \\
&-\ii\iek{n_n(t)+n_l(t)}\fbraket{\tilde l}{\Ge}{\tilde n} \\
&+2 n_n(t)  \sum_{ijk} 
 \tilde \gamma_{2,ijnk}(t) \fbraket{\tilde l \tilde k}{\Vee}{\tilde i \tilde j} \\
 &-\bigl[2 n_l(t)\sum_{ijk} \tilde \gamma_{2,ijlk}(t) \fbraket{\tilde n \tilde k}{\Vee}{\tilde i \tilde j}\bigr]^\ast \\
 &-4\ii n_n(t) n_l(t)  \sum_{jk}  \tilde \gamma_{2,ljnk}(t) \fbraket{\tilde k}{\Ge}{\tilde j}.
 \label{eq:mat_dot_gamma1}
\end{split}
\end{align}
Multiplying
\begin{align}
 \dot{\hat\gamma}_1(t)&=\sum_k \ketbra{\dot{\tilde k}}{\tilde k}+\sum_k\ketbra{\tilde k}{\dot{\tilde k}}
 \label{eq:gamma1_dot}
\end{align}
 from the right with $\ket{\tilde n}$ and rearranging terms one obtains
\begin{align}
n_n(t) \ket{\dot{\tilde n}} =\dot{\hat\gamma}_1(t)\ket{\tilde{n}}-\braket{\dot{\tilde n}}{\tilde n}\ket{\tilde n}-\sum_{k\neq n} \braket{\dot{\tilde 
k}}{\tilde n}\ket{\tilde k}.
\label{eq:A2}
\end{align}
Using the time derivative of the orthogonality relation $\braket{\tilde l 
(t)}{\tilde 
n (t)} =\delta_{l,n}n_n(t)$,
\begin{equation}
 \braket{\dot{\tilde l}}{\tilde n}+\braket{\tilde l}{\dot{\tilde n}} =\delta_{l,n}\dot n_n(t),
\label{eq:A6}
\end{equation}
allows us to rewrite  \eqref{eq:A2} as
\begin{align}
\ket{\dot{\tilde n}} =\dot{\hat\gamma}_1(t)\ket{\tilde{n}}+\braket{{\tilde n}}{\dot{\tilde n}}\ket{\tilde n}-\dot n_n\ket{\tilde n}+\sum_{k\neq n} \braket{{\tilde 
k}}{\dot{\tilde n}}\ket{\tilde k},
\label{eq:A2a}
\end{align}
which is still implicit. However, multiplying  \eqref{eq:A2a} from the left with $\bra{\tilde l}$ for $\bra{\tilde l}=\bra{\tilde n}$
\begin{equation}
\begin{split}
 \dot n_n(t)&=\frac{1}{n_n(t)}\fbraket{\tilde n}{\dot {\hat\gamma}_1(t)}{\tilde n}=4 \Im\sum_{ijl} \tilde \gamma_{2,ijnl}(t) \fbraket{\tilde n \tilde 
l}{\Vee}{\tilde i 
\tilde j}\\
&- 2\fbraket{\tilde n}{\Ge}{\tilde  n}
 -4 n_n(t)  \sum_{jl} \tilde \gamma_{2,njnl}(t) \fbraket{\tilde l}{\Ge}{\tilde j}
\label{eq:A3}
\end{split}
\end{equation}
results, for $n_l(t)\neq n_n(t)$ at time $t$
\begin{equation}
  \braket{\tilde l}{\dot{\tilde n}}=-\braket{\dot {\tilde l}}{{\tilde n}}=\frac{\fbraket{\tilde l}{\dot {\hat\gamma}_1(t)}{\tilde n}}{n_n(t)-n_l(t)},
\label{eq:A4}
\end{equation}
and for $n_l(t)=n_n(t)$ at time $t$ if $l\neq n$
\begin{equation}
 \fbraket{\tilde l}{\dot {\hat\gamma}_1(t)}{\tilde n}=0.
\end{equation}
The last expression, with  \eqref{eq:mat_dot_gamma1}, yields a conservation rule at time $t$ for $n_l(t)=n_n(t)$ and $\braket{ l}{ n}=0$:
\begin{equation}
\begin{split}
   &\sum_{ijk} \tilde \gamma_{2, ijnk}(t) \fbraket{\tilde l \tilde k}{\Vee}{\tilde i \tilde{j}}-2\ii\fbraket{\tilde l}{\Ge}{\tilde n}
\\&= \sum_{ijk} 
\tilde \gamma_{2, lkij}(t) \fbraket{\tilde i \tilde j}{\Vee}{\tilde n \tilde{k}}\\
&+4\ii  n_n(t) \sum_{jk}\tilde \gamma_{2, ljnk}(t) \fbraket{\tilde k}{\Ge}{\tilde{j}}.
\end{split}
\end{equation}
The condition $n_l(t)=n_n(t)$ for $\braket{l}{n}=0$ means that these 
orbitals at time $t$ are degenerate, and one may choose any orthogonal pair from the subspace they span. 

The term $\braket{{\tilde n}}{\dot{\tilde n}}$ depends on the phase choice for the RNOs. In principle one can use any value, keeping in mind that  $\tilde \gamma_{2, ijnk}(t)$ will depend on the choice. For the phase  convention introduced in \cite{tdrnot2} (with slight modifications to ensure that during imaginary time propagation the norms of RNOs $\ket{\tilde n}$ and $\ket{\tilde n\pr}$ remain the same, see Appendix~\ref{sec:phase_conv}) 
\begin{equation}
\begin{split}
    \ii\braket{{\tilde n}}{\dot{\tilde n}}&=\frac{1}{2} \fbraket{\tilde n}{\hat{h}}{\tilde n}+\frac{1}{2} \fbraket{\tilde n\pr}{\hat{h}}{\tilde n\pr} \\&+\Re\sum_{ijk} 
\tilde \gamma_{2, ij nk}\fbraket{\tilde n\tilde k}{\Vee}{\tilde i\tilde j}+\ii\frac{\dot n_n}{2}.
\label{eq:alpha}
\end{split}
\end{equation}
Inserting \eqref{eq:gamma_1}, \eqref{eq:alpha}, \eqref{eq:A3}, and \eqref{eq:A4} with \eqref{eq:mat_dot_gamma1} into \eqref{eq:A2a} one obtains 
 the explicit EOM for the RNOs \eqref{eq:eom-rno}--\eqref{eq:eom-end}.

Note that we can use \eqref{eq:Bnk1} to calculate $\mat{B}_{nk}(t)$ only if at time $t$ we have $n_k(t) \neq n_n(t)$. Otherwise 
\begin{align}
\begin{split}
 &{\mat{B}}_{nk}(t)
    =
     \frac{\ii}{n_n(t)}\braket{\tilde{k}}{\dot{\tilde n}}- \frac{
\fbraket{\tilde k}{h(t)}{\tilde n}}{n_n(t)}\\
&-2 \sum_{ijl}
\frac{\tilde \gamma_{2, kijl}(t) \fbraket{\tilde j \tilde
l}{\Vee}{\tilde n \tilde{i}}}{n_n(t)} -\ii\frac{ \fbraket{\tilde
k}{\Ge}{\tilde
n}}{n_n(t)}\\
&-4\ii\sum_{jl}\tilde \gamma_{2, kjnl}(t) \fbraket{\tilde
l}{\Ge}{\tilde{j}}
\label{eq:Bnk_deg}
\end{split}
\end{align}
if $ n_k(t) = n_n(t)$.
Without imaginary potential it is always possible to choose such linear combination that 
$ {\mat{B}}_{nk}(t)
    =0$.
However, with imaginary potential the situation is more complicated. For two-electron systems  one can still set 
$ {\mat{B}}_{nn'}(t)
    =0 $
but in general it is  advisable to use 
\eqref{eq:Bnk_deg} with some value for $\braket{\tilde{k}}{\dot{\tilde n}}$ which does not change $\gamma_{2, ijnk}(t)$, e.g., 0.
%

\section{RNO phase convention and proof of  \eqref{eq:alpha}}
\label{sec:phase_conv}
The two-electron state within a particular phase convention for the RNOs can be written as \cite{tdrnot2}
\begin{equation}
  \ket{\fullwf(t)}=\sum_{i \odd} \tilde g_i(t) \kiek{\ket{\tilde i \tilde i'} -\ket{\tilde i' \tilde i}}, \quad \tilde 
g_i(t)=\frac{\ee^{\ii\varphi_i}}{\sqrt{2n_i(t)}}.
\label{eq:b1}
\end{equation}
The phase factors $\ee^{\ii\varphi_i}$ for the helium singlet can be chosen 
\begin{equation}
  \ee^{\ii\varphi_i} = 2\delta_{k,1} -1, \quad k \text{ } \odd,
\end{equation}
\begin{widetext}
\noindent leading to real expansion coefficients 
\begin{equation}
  \tilde \gamma_{2, ijkl}(t)=\tilde{g}_i(t)\tilde{g}_k(t)\delta_{i,j'}\delta_{k,l'} =\iek{-1}^{i-k}\frac{\ee^{\ii\kiek{\varphi_i-\varphi_k}}}{2\sqrt{n_i(t)n_k(t)}}\delta_{i,j'}\delta_{k,l'}.
  \label{eq:gamma2ijkl}
\end{equation}
Inserting \eqref{eq:b1} into the left hand side of the TDSE \eqref{eq:TDSE} leads to
\begin{equation}
 \hat H^{(1, 2)}(t)\ket{\fullwf (t)}=\ii \sum_{i \odd}  \dot{\tilde g}_i(t) \kiek{\ket{\tilde i \tilde i'} -\ket{\tilde i' 
\tilde i}} +\ii \sum_{i \odd}{\tilde g}_i(t) \kiek{\ket{ \dot{\tilde i} \tilde i'} + \ket{\tilde i \dot{\tilde i}'} - \ket{\dot{\tilde i}' \tilde 
i} - \ket{\tilde i' \dot{\tilde i}}} .
\end{equation}
Multiplying from the left by $2 \tilde{g}_k(t)\bra{\tilde k \tilde k\pr}$ for an odd $k$ and making use of $2\tilde g^2_k(t) n_k(t)=1$ gives
\begin{equation}
2 \tilde{g}_k(t)\fbraket{\tilde k\tilde k\pr}{\hat H(t)}{\fullwf (t)}
 = \ii  \frac{\dot{\tilde g}_k(t)}{{\tilde g}_k(t)}n_k(t)+\ii \iek{\braket{\tilde 
k}{\dot{\tilde k}} +\braket{\tilde k\pr}{\dot{\tilde k}\pr} } =\ii \iek{\braket{\tilde 
k}{\dot{\tilde k}} +\braket{\tilde k\pr}{\dot{\tilde k}\pr} -\frac{\dot{n}_k(t)}{2}} .
\label{eq:b6}
\end{equation}
Inserting \eqref{eq:b1} into the right hand side of \eqref{eq:b6} gives
\begin{equation}
 2 \tilde{g}_k(t)\fbraket{\tilde k\tilde k\pr}{\hat H(t)}{\fullwf (t)} =\fbraket{\tilde k }{\hat h(t) - \ii \Ge}{\tilde k}+\fbraket{\tilde k' }{\hat h(t) - \ii \Ge}{\tilde k'} +2\sum_{i\odd} \tilde g_i(t) \tilde{g}_k(t) \kiek{\fbraket{\tilde k \tilde k \pr}{\Vee}{\tilde i \tilde i \pr}+\fbraket{\tilde k \tilde k \pr}{\Vee}{\tilde i\pr 
\tilde i }}, 
\label{eq:b7}
\end{equation}
which, using \eqref{eq:A3} and \eqref{eq:gamma2ijkl}, simplifies to
\begin{equation}
2 \tilde{g}_k(t)\fbraket{\tilde k\tilde k\pr}{\hat H(t)}{\fullwf (t)} =
\fbraket{\tilde k }{\hat h(t)}{\tilde k}+\fbraket{\tilde k' }{\hat
h(t)}{\tilde k'} +2\Re\sum_{ijl} \tilde \gamma_{2, ij kl}\fbraket{\tilde
k\tilde l}{\Vee}{\tilde i\tilde j} + \ii\frac{\dot{n}_k(t)}{2}.
\label{eq:b8}
\end{equation}
Combining \eqref{eq:b6} and  \eqref{eq:b8} one arrives at
\begin{equation}
 \ii \braket{\tilde n}{\dot{\tilde n}}+\ii \braket{\tilde n \pr}{\dot{\tilde n}\pr}= \fbraket{\tilde n}{\hat{h}(t)}{\tilde n}+ 
\fbraket{\tilde n\pr}{\hat{h}(t)}{\tilde n\pr} +2\Re\sum_{ijl} \tilde \gamma_{2, ij nl}\fbraket{\tilde n\tilde l}{\Vee}{\tilde i\tilde j} + \ii{\dot{n}_n(t)}.
\end{equation}
There is the freedom to distribute the right hand side between $\ii \braket{\tilde n}{\dot{\tilde n}}$ and $\ii \braket{\tilde n \pr}{\dot{\tilde n}\pr}$.
We deviate slightly from the choice in  \cite{tdrnot2} and set 
\begin{equation}
\begin{split}
 \ii \braket{\tilde n}{\dot{\tilde n}}= \ii \braket{\tilde n\pr}{\dot{\tilde n}\pr}= &\frac{1}{2}\fbraket{\tilde n}{\hat{h}(t)}{\tilde n}+\frac{1}{2}\fbraket{\tilde n\pr}{\hat{h}(t)}{\tilde n\pr}+\Re\sum_{ijl} \tilde\gamma_{2, ij nl}\fbraket{\tilde n\tilde l}{\Vee}{\tilde i\tilde j} + \ii 
\frac{\dot{n}_n(t)}{2} , 
\end{split}
\end{equation}
which has the advantage that the OCs of RNOs $\ket{n}$ and $\ket{n'}$ remain automatically equal during imaginary time propagation. 
\end{widetext}


%

\end{document}